\documentclass[aps,prl,preprint,nopacs,superscriptaddress]{revtex4}
\usepackage{graphicx}
\usepackage{verbatim}
\usepackage{mathrsfs}
\usepackage{amsmath,amsfonts,amssymb}
\usepackage{graphicx}

\newcommand{\ud}{\mathrm{d}}

\def\3{2.8in}    %used for figure widths
\def\2{2.5in}
\def\4{3.0in}

\def \beq {\begin{equation}}
\def \eeq {\end{equation}}
\begin{document}

\title{Mirror protected Dirac fermions on Weyl semimetal NbP surface}

\author{Hao~Zheng\footnote{These authors contributed equally to this work.}}
\affiliation {Department of Physics, Princeton University, Princeton, New Jersey 08544, USA}

\author{Guoqing~Chang$^*$}
\affiliation {Centre for Advanced 2D Materials and Graphene Research Centre National University of Singapore, 6 Science Drive 2, Singapore 117546}
\affiliation {Department of Physics, National University of Singapore, 2 Science Drive 3, Singapore 117542}

\author{Shin-Ming~Huang$^*$}
\affiliation {Department of Physics, National Sun Yat-sen University, Kaohsiung 80424, Taiwan}

\author{Cheng~Guo}
\affiliation {International Center for Quantum Materials, School of Physics, Peking University, China}

\author{Xiao~Zhang}
\affiliation {International Center for Quantum Materials, School of Physics, Peking University, China}

\author{Songtian~Zhang}
\affiliation {Department of Physics, Princeton University, Princeton, New Jersey 08544, USA}

\author{Jiaxin~Yin}
\affiliation {Department of Physics, Princeton University, Princeton, New Jersey 08544, USA}

\author{Su-Yang~Xu}
\affiliation {Department of Physics, Princeton University, Princeton, New Jersey 08544, USA}

\author{Ilya~Belopolski}
\affiliation {Department of Physics, Princeton University, Princeton, New Jersey 08544, USA}

\author{Nasser~Alidoust}
\affiliation {Department of Physics, Princeton University, Princeton, New Jersey 08544, USA}

\author{Daniel~S.~Sanchez }
\affiliation {Department of Physics, Princeton University, Princeton, New Jersey 08544, USA}

\author{Guang~Bian}
\affiliation {Department of Physics, Princeton University, Princeton, New Jersey 08544, USA}

\author{Tay-Rong Chang}
\affiliation{Department of Physics, National Tsing Hua University, Hsinchu 30013, Taiwan}

\author{Titus Neupert}
\affiliation{Department of Physics, University of Zurich, Winterthurerstrasse 190, 8057 Zurich, Switzerland}

\author{Horng-Tay Jeng}
\affiliation{Department of Physics, National Tsing Hua University, Hsinchu 30013, Taiwan}

\author{Shuang~Jia}
\affiliation {International Center for Quantum Materials, School of Physics, Peking University, China}
\affiliation {Collaborative Innovation Center of Quantum Matter, Beijing,100871, China}

\author{Hsin~Lin}
\affiliation {Centre for Advanced 2D Materials and Graphene Research Centre National University of Singapore, 6 Science Drive 2, Singapore 117546}
\affiliation {Department of Physics, National University of Singapore, 2 Science Drive 3, Singapore 117542}

\author{M. Zahid~Hasan\footnote{mzhasan@princeton.edu}}
\affiliation {Department of Physics, Princeton University, Princeton, New Jersey 08544, USA}


\begin{abstract}
The first Weyl semimetal was recently discovered in the NbP class of compounds. 
Although the topology of these novel materials has been identified, the surface properties are not yet fully understood. 
By means of scanning tunneling spectroscopy, we find that NbP’s (001) surface hosts a pair of Dirac cones protected by mirror symmetry. 
Through our high resolution spectroscopic measurements, we resolve the quantum interference patterns arising from these novel Dirac fermions, and reveal their electronic structure, including the linear dispersions. 
Our data, in agreement with our theoretical calculations, uncover further interesting features of the Weyl semimetal NbP’s already exotic surface. 
Moreover, we discuss the similarities and distinctions between the Dirac fermions here and those in topological crystalline insulators in terms of symmetry protection and topology.

\end{abstract}
\date{\today}
\maketitle
The recent discovery of the Weyl semimetal state in the TaAs class of compounds (TaAs, TaP, NbAs and NbP) has received significant attention worldwide\cite{Weyl1, Weyl2, TaAs1, TaAs2}. 
The existence of Weyl fermions of opposite chiralities in the bulk of these materials gives rise to a new type of surface state, the Fermi arcs, which were once considered as the only topological response on a Weyl semimetal surface.
However,  the possibility of other novel surface states was very recently reported in theoretical works, such as the helicoid surface state and the time-reversal symmetry protected surface Dirac cone \cite{Fang, Lau}, which have yet to be experimentally verified. 
This clearly indicates that the boundary state of three-dimensional topological semimetals is not yet fully understood.

In this paper, we theoretically predict the existence of previously overlooked surface Dirac cones on TaAs family compounds by first-principle calculations. 
Next, we choose NbP as a platform on which to perform our experiments. 
Among the four compounds in the TaAs class, NbP has the weakest spin-orbit coupling (SOC), which leads to a relatively small separation between the Weyl nodes\cite{TaAs1, TaAs2, NbAs, TaP, NbP1, NbP2, NbP3}. 
This has hindered a direct and unambiguous observation of the Fermi arcs in this compound based on the currently available experimental resolution. 
However, we show that the weak SOC approximately realizes another type of novel surface state, \textit{i.e.}, mirror symmetry protected massless Dirac quasi-particle without SOC.
Historically, Dirac surface states have been observed in the $\mathbb{Z}_2$ topological insulators such as Bi$_2$Se$_3$ and the topological crystalline insulators (TCIs) such as SnTe \cite{TI, TCI1, TCI2, TCI3}. In all these topological materials, SOC not only plays an essential role in the formation of the respective topological states, but also directly leads to the spin-momentum locking of the Dirac surface states. Dirac surface states without SOC have been proposed in some novel types of TCIs (different from the SnTe type) but have not been realized in experiment~\cite{Fu, TCI4}.
In NbP the Dirac cones are located at about 300~meV above the Fermi level, and therefore cannot be accessed by conventional photoemission measurements. 
We employ scanning tunneling spectroscopy (STS) to overcome this obstacle and experimentally verify the prediction.

The experiments are performed on a Unisoku ultra high vacuum system, which contains a low-temperature scanning tunneling microscope (STM). Our high quality samples are cleaved at liquid nitrogen temperature and measured at 4.6~K. The ${\ud}I/{\ud}V$ signals are obtained from a lock-in amplifier with modulation voltages at 1 to 8~mV. First-principle calculations \cite{Ab1, Ab2, Ab3} are adopted to simulate the band structure and quasi-particle interference (QPI) patterns (see more details in \cite{QPI1, QPI2}).  

Sharing the same space group and similar lattice constants, all members of the TaAs class of Weyl semimetals possess similar electronic band structures and Fermi surfaces. In the lower panel of Fig.~1(a), we schematically draw the Fermi surface on the (001) surface of TaAs, TaP, NbAs, and NbP. The surface band contours can be divided into three groups according to their different line shapes. Namely, four tadpole-shaped contours [pink lines in Fig.~1(a)] are located along the $\bar{X}$-$\bar{M}$ and $\bar{Y}$-$\bar{M}$ directions, which include the exotic Fermi arcs as parts of the contours. In addition, two elliptical contours and two bowtie-shaped contours sit in the vicinity of the $\bar{Y}$ and $\bar{X}$ points respectively, which are considered to be trivial pockets as they are irrelevant to the Weyl fermions (cones). However, our calculation, without taking SOC into account, shows that the bowtie-shaped contours themselves feature two-dimensional massless Dirac quasi-particles (Figure S2 shows the massive case with considering SOC) which are protected by the $\bar{X}$-$\bar{M}$-$\bar{X}$ mirror plane. As displayed in the upper panel of Fig.~1(a), a pair of surface Dirac cones is located along the $\bar{M}$-$\bar{X}$-$\bar{M}$ line, at an energy above the Fermi level. As the energy is varied towards the Fermi level, the two Dirac cones expand in size, eventually merging into each other and evolving into a bowtie-shaped contour. 

As the predicted massless Dirac fermions appear in a SOC-vanishing crystal, we decide to choose NbP to verify our calculated results. 
Figure~1(b) shows the energy-momentum ($E$-$k$) dispersion of the surface energy bands along the $\bar{X}$-$\bar{M}$-$\bar{Y}$ high symmetry line of the NbP(001) surface. One can clearly discern that a Dirac node forms in the unoccupied band structure. We then carry out STM measurements on our single-crystalline NbP samples. In Fig.~1(c), the constant-current STM topographic image demonstrates a highly ordered square lattice with a measured lattice constant of 3.4 {\AA}, proving that our NbP crystal indeed cleaves at the (001) plane. A typical tunneling conductance (${\ud}I/{\ud}V$) spectrum is represented in Fig.~1(d). It reveals a non-vanishing conductance at the Fermi level, which confirms the (semi-)metallic nature of our NbP sample, and more importantly, has a minimum at about 300 meV above Fermi level, which indicates the energy position of the predicted Dirac node.

QPI patterns arise from surface standing waves, which are induced by surface scatterings at point defects, and have become a common approach for detecting the surface electronic structure of a crystal \cite{STM1, STM2, STM3, STM4, STM5, STM6}. This method has been applied to probe the TaAs surface in both occupied and unoccupied states \cite{QPI3, QPI4, QPI5}. QPI patterns are obtained by a Fourier transformation (FT) of a real space ${\ud}I/{\ud}V$ map. In Fig.~2(a), we present an example of such a FT-${\ud}I/{\ud}V$ map measured on the NbP (001) surface. One can clearly observe that the QPI pattern (FT-${\ud}I/{\ud}V$ map) contains rich information about the electronic structure. Benefiting from our high resolution measurements, we are able to clearly discern line-shaped contours in the QPI pattern. Our current experimental QPI map is consistent with our previous results \cite{QPI1, QPI2}. We can thus confirm that the cleaved surface here is the P-terminated (001) surface, and unambiguously assign all of the QPI contours  to scattering processes between the surface Fermi pockets. We can see that the vertical bowtie-shaped QPI contour at the center of the map arises from the intra-pocket scattering vectors inside the bowtie-shaped contours at the $\bar{X}$ points in the Fermi surface [in Fig.1(a)]. The horizontal ellipse, which is perpendicular to the bowtie contour, originates from the intra-pocket scattering in the elliptical contours at the $\bar{Y}$ points. The four nearly square-shaped contours are induced by inter-pockets scattering between the bowties and ellipses. 
One may notice some features located near the Bragg points [marked as ''B" in Fig.~2(a)], which are simply the replicas of the central QPI features.    

Having verified the existence of the predicted Dirac cones in the bowtie-shaped contours in the surface electronic structure, we now turn our attention to the corresponding bowtie-shaped QPI pockets in the FT-${\ud}I/{\ud}V$ maps. Figure~2(d) shows the enlarged voltage-dependent images of the the bowtie-shaped QPI contours, clearly demonstrating the evolution of the interesting QPI features. In the energy range from -100~meV to 40~meV, the QPI contours show obvious bowtie shapes, which shrink in size with increasing energies. At 125~meV, the QPI contour breaks into three parts: a central feature and two cobble-shaped contours above and below the central part. This transition of the QPI feature in $Q$-space serves as explicit evidence of a Lifshitz transition in the band structure in $k$-space, indicating that the two Dirac cones separate from each other at this energy. Focusing on the two cobble-shaped QPI contours, one can notice that their diameters gradually decrease with increasing energies and collapse into two large dots at 275~meV. When the energy is increased to 400~meV, only the central QPI feature remains. Having resolved the contour shapes in several successive QPI patterns, we now focus our study on the energy-scattering vector ($E$-$Q$) dispersion relationship of the predicted Dirac cones. Figure~2(b) presents a line cut plot taken along the dotted line in Fig. 2(a), which crosses the center of one cobble-shaped QPI contour. One can observe a distinctive $\Lambda$-shaped $E$-$Q$ dispersion, indicating the linearly dispersing Dirac cone, with the Dirac node at about 300~meV in the unoccupied surface band.  In order to further prove the existence of the predicted Dirac cone, we take another line cut in the orthogonal direction and present it in Fig. 2 (c). We note that the line crosses the central part of the QPI map, in which all of the small Q scattering vectors mix up and complicate the central feature. Consequently, Fig. 2(c) is not as clean as Fig. 2(b). Nevertheless, away from this central region, we can still resolve the Dirac node and one branch of the cone.

In order to gain further insights into the surface Dirac cones, we carry out comprehensive \textit{ab-initio} calculations (without SOC) of the electronic structure and QPI patterns of NbP(001). 
Figure~3(a) shows the calculated surface constant energy contours (CEC) in the first surface Brillouin zone (BZ) at 250~meV, which is slightly below the energy of the Dirac nodes. 
At this energy, apart from the two pairs of Dirac cones, the remaining contours all almost entirely disappear. 
Each pair of Dirac cones sits on the zone boundary and extends beyond the first BZ. 
In Fig.~3(b), we present the complete image of one pair of Dirac cones. 
They manifest themselves as cobble-shaped contours, indicating the anisotropic Fermi velocities along the $k_x$ and $k_y$ directions. 
Based on the CEC, we further simulate the QPI pattern through a joint-density-of-state calculation, shown in Fig.~3(c). 
In Fig.~3(d), we enlarge the features of interest and explicitly resolve two cobble-shaped  QPI contours which display almost exactly the same shapes as shown in the CEC [Fig.~3(b)]. 
We identify the dominant scattering vector $Q$ which links the two Dirac cones in Fig.~3(b), and find that this scattering produces the cobble-shaped QPI pocket in Fig.~3(d). 
The simulated QPI pattern reproduces our measurement in Fig.~3(e) very well. 
In addition to the qualitative agreement, we also quantitatively compare our theoretical and experimental results. We find that the $k_y$ coordinates of the Dirac cones are $k_y \cong \pm 0.142 (2\pi/a)$ from experiment, while the theoretical values are $k_y = \pm 0.136 (2\pi/a)$. 
Additionally, since the real material indeed posses finite SOC, we also performed the calculations of CEC and QPI patterns of NbP (001) by taking SOC into account (Fig.S3). 
The results with (Fig. 3) and without (Fig. S3) SOC does not display obvious difference. 
Taking into account all the above evidence, we firmly establish the existence of two-dimensional Dirac-type surface states on the NbP(001) surface. 

Finally, we discuss the nature of our novel surface Dirac cones. A surface Dirac fermion can be described by the low-energy Hamiltonian $H(\mathbf{q})=v_{x}q_{x}\sigma_{x}+v_{y}q_{y}\sigma_{y}$, where $\mathbf{q}$ denotes the relative momentum to the Dirac node and $\sigma_{x,y}$ are Pauli matrices. To preserve the gapless Dirac node, a symmetry constraint is required, or a perturbation on the third Pauli matrix will easily open a gap. For this reason, they are found in time-reversal-symmetric topological insulators or in mirror-symmetric TCIs. In our case, the Dirac nodes are protected by mirror symmetry, because the two bands which cross to form the Dirac cone have opposite mirror eigenvalues. [If we artificially break the mirror symmetry, the Dirac cone will gap out (see details in Fig. S2 and supplementary text).] However, unlike TCIs, they are not guaranteed to exist due to a mirror Chern number. This is because the 2D space group (\textit{pmm}) in a NbP(001) surface has vanishing mirror Chern numbers for spinless systems \cite{Dong2016}. Interestingly, the Dirac nodes on our NbP surfaces are movable on a mirror-symmetric line as they are locally protected by the mirror symmetry. Furthermore, they are also able to move to the other mirror-symmetric lines, circulating around $\bar{\Gamma}$-$\bar{X}$-$\bar{M}$-$\bar{Y}$-$\bar{\Gamma}$, which is not possible on the surface of a TCI. During the motion, a Dirac cone will cross a time-reversal-invariant momentum (TRIM) point before turning into another mirror-symmetric line. Contrary to the case in TCIs, the Dirac cone on NbP does not collide and annihilate with the other. It is a unique property of the NbP surface spinless Dirac cone. In this spinless case, the Dirac node cannot be destroyed at the intersections of distinct mirror-symmetric lines, \textit{i.e.} TRIMs. This is unlike the case of spinful TCIs, in which the the number of Dirac nodes on each mirror line is required by the mirror Chern number, and Dirac nodes at TRIMs are protected by time-reversal symmetry. We provide a Hamiltonian to explain the phenomenon, 
\beq
H(\mathbf{k})=A(k_{x}^2 + \nu k_{y}^2-\lambda)\sigma_{x}+Bk_{x}k_{y}\sigma_{z},
\eeq
where $\mathbf{k}$ is relative to a TRIM, such as $\bar{X}$, and $A, \,B$, $\nu$, $\lambda$ are real parameters. We introduce $\lambda$ to model a perturbation for shifting the Dirac fermions. In this basis, the mirror operators are given by $M_{x}=M_{y}=\sigma_{x}$ (or $M_{y}=-M_{x}$), such that the Hamiltonian satisfies $M_{x(y)}H(k_{x},k_{y})M_{x(y)}^{-1}=H(-k_{x},k_{y})=H(k_{x},-k_{y})$. We set $\nu<0$. By tuning $\lambda$ from positive to negative, the Dirac nodes are shifted from $(k_{x},k_{y})=(\pm \sqrt{\lambda},0)$ to $(k_{x},k_{y})=(0,\pm \sqrt{-\lambda})$. When two Dirac fermions meet at $\bar{X}$ ($\lambda=0$), they hybridize into a quadratic Dirac fermion, which is stable as it is protected by both $M_{x}$ and $M_{y}$.
The scenario where $\nu>0$ and two Dirac nodes appear around $\bar{X}$ when $\lambda>0$ was not observed in this system.

In summary, we employ first-principle simulations and STS measurements to investigate the P-terminated (001)-surface electronic structure in the Weyl semimetal NbP. 
In addition to the previously observed large magnetoresistance and ultra high mobility \cite{MR1,MR2}, our observations here not only discover the first Dirac surface quasiparticle without SOC, but also highlight further exotic properties in NbP in that it hosts both Weyl and Dirac type excitation modes. 
Remarkably, the mobility of these spinless Dirac nodes in $k$-space is not restricted, in sharp contrast to the behavior found in spinful TCIs.
The mirror symmetry protected Dirac cone discovered here may lead to potential applications in topological devices.

The STM measurements at Princeton University were supported by the Gordon and Betty Moore Foundations EPiQS Initiative through grant GBMF4547 (Hasan). H.Z also thanks the support from National Natural Science Foundation of China (No.11674226). Theoretical calculations at National University of Singapore were supported by the National Research Foundation, Prime Minister's Office, Singapore, under its NRF fellowship (NRF Award no. NRF-NRFF2013-03). NbP crystal growth was supported by National Basic Research Program of China (grant nos. 2013CB921901 and 2014CB239302) T.-R. C and H.-T. J. are supported by National Science Council, Academia Sinica, and National Tsing Hua University, Taiwan, and also thank NCHC, CINC-NTU, and NCTS, Taiwan for technical support.

\newpage

\clearpage
\begin{figure*}
\centering
\includegraphics[width=8cm]{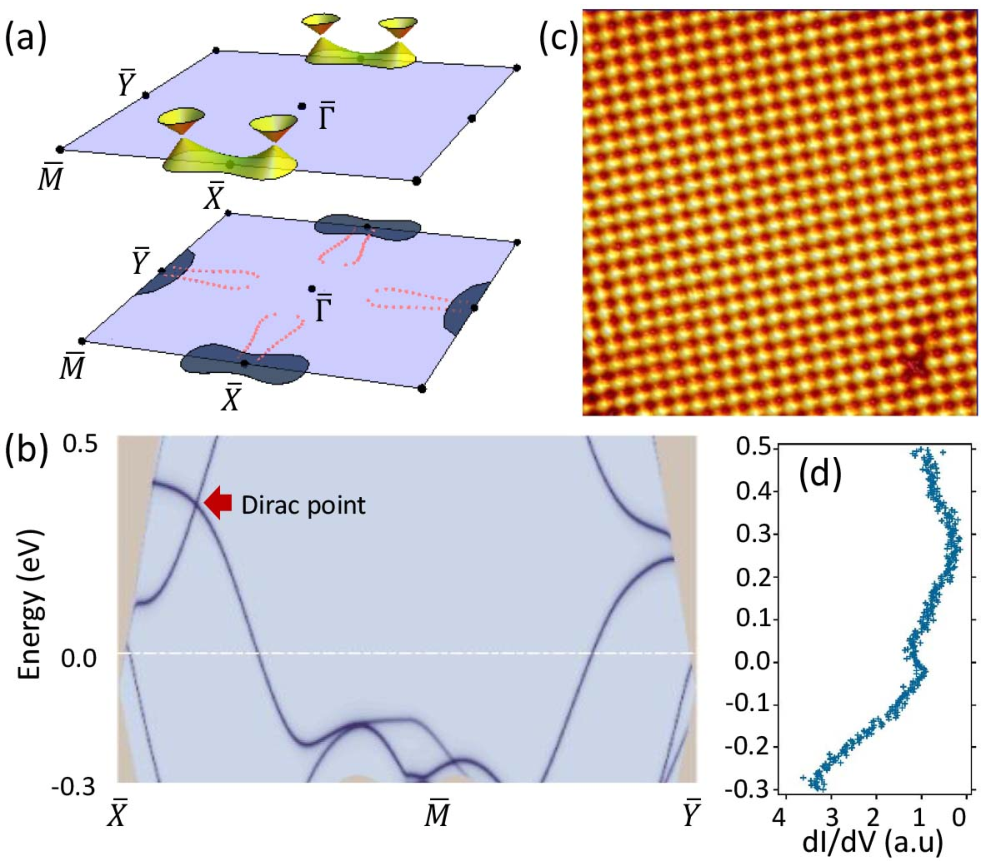}

\caption{(a) Illustrations of the two-dimensional Dirac cones (upper panel) and the Fermi surface (lower panel) in the first surface Brillouin zone (BZ) of NbP(001).
High symmetry points are marked on the sketches.
Red dotted lines are the Fermi arc derived surface contours.
Dark blue regions are the surface states, which are ''trivial" in the context of Weyl physics, but feature interesting Dirac cone type electronic structures.
(b) The surface electronic band structure along $\bar{X}$-$\bar{M}$-$\bar{Y}$.
Two bands cross each other and form a Dirac node, indicated by the arrow.
(c) A constant-current STM image on our high-quality NbP crystal's (001) surface, showing the atomically ordered lattice with only one point defect in the $7.7 \times 7.7$~nm$^2$ area. 
(d) A typical d\textit{I}/d\textit{V} spectrum on NbP(001) surface far from defects.
}
\end{figure*}

\clearpage
\begin{figure*}
\centering
\includegraphics[width=13cm]{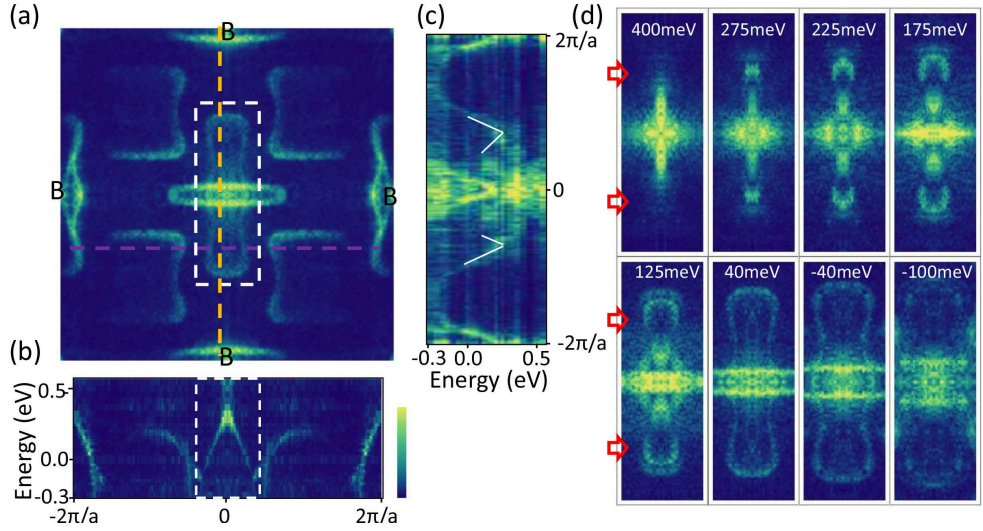}
\caption{
(a) A Fourier-transformed d\textit{I}/d\textit{V} map (50~mV) representing the quasi-particle interference (QPI) pattern on the NbP(001) surface.
The four Bragg points are indicated by B.
(b) and (c) Energy-resolved QPI features along orthogonal directions taken along the purple and yellow dashed lines in (a) respectively.
A Dirac cone shaped feature can be seen inside the white dashed rectangle.
White lines in (c) serve as guides to the eye for the Dirac cones. 
(d) A series of QPI patterns at indicated energies,  clipped from the white dashed rectangle in (a).
Arrows indicate the positions of Dirac nodes.
}
\end{figure*}

\clearpage
\begin{figure*}
\centering
\includegraphics[width=7cm]{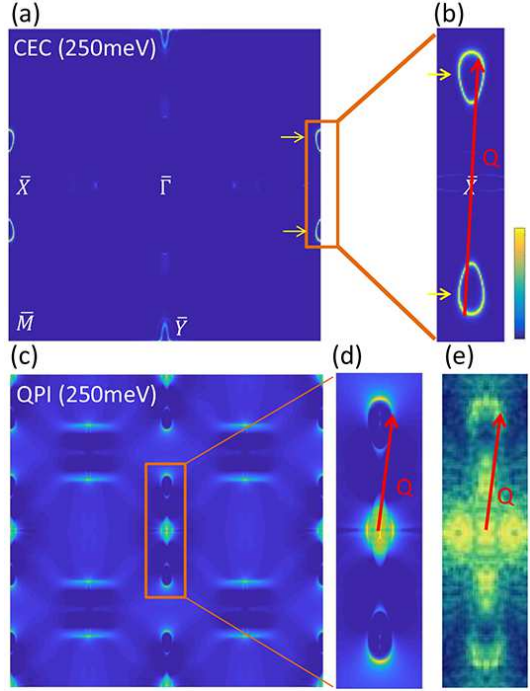}
\caption{
(a) The theoretically calculated constant energy contour (CEC) at 250~meV above the Fermi level.
High symmetry points in the first Brillouin zone are marked.
The positions of the Dirac cones are indicated by the arrows.
(b) A zoomed-in view of the area marked by the rectangle in (a), where a pair of Dirac cones are visible. 
A dominant scattering vector $Q$ links the two Dirac contours.
(c) A numerically simulated QPI pattern at the same energy as (a).
(d) The zoomed-in view of the central region of (c), indicated by a red rectangle.
(e) The experimentally acquired QPI pattern. 
Its main features are well reproduced by our simulation in (d). The scattering vector $Q$ in (d) and (e) are same as in (b).
}
\end{figure*}

\end{document}